\documentclass[10pt,a4paper]{article}
\usepackage[utf8]{inputenc}
\usepackage[english,czech]{babel}
\usepackage[T1]{fontenc}
\usepackage{amsmath}
\usepackage{amsfonts}
\usepackage{amssymb}
\usepackage{graphicx}
\usepackage[a4paper]{geometry}

\begin{document}

\author{Jakub Kakona $^1$, Pavel Kovar $^1$, Martin Kakona $^2$\\
\\
$^1$ Dept. of Radio Engineering,\\ 
Czech Technical University in Prague,\\
Technicka 2,\\
166 27 Praha 6, \\
Czech Republic \\
\\
$^2$ Dept. of Dosimetry and Application of Ionizing Radiation,\\
Czech Technical University in Prague, \\
Brehova 7,\\
115 19 Praha 1,\\
Czech Republic
\texttt{{kakonjak, kovar}@fel.cvut.cz}}
\title{Bolidozor - Distributed radio meteor detection system}

\maketitle

\begin{abstract}
Most of the meteor radioastronomical radars are backscatter radars which cover only a small area of the atmosphere. Therefore a daytime meteor flux models are based on sparse data collected by only a few radar systems. To solve this issue, a radar system with a wide coverage is required. We present a new approach of open-source multi-static radio meteor detection system which could be distributed over a large area. This feature allows us to detect meteor events taking place over a larger area as well and gather more uniform data about meteor flux and possibly about meteor trajectories. Therefore, a new network which detects atmospheric meteor impacts and produce scientifically valuable data about each detected event is needed. 
\end{abstract}



\section{Introduction}

Meteors, the products of meteoroids travelling in space, are studied for many decades. Results of such research help us to understand the evolution of not only the planetary system but the interplanetary medium \cite{interplanetary_medium} as well. As the density of the interplanetary medium is low, large statistical and long term continuous data set is necessary to describe its properties. 
 
\subsection{Optical observation methods}

Historically, the observations were based mainly on optical methods such as camera fireball network \cite{Bland01102004}.
Results of these observations were not  continuous, a fact which was caused by a non-uniform coverage and sensitivity of detection systems. There are several reasons for data absence, one of them being a weather dependency of the observation method. It results from the fact that optical method needs mainly clear air and low light background \cite{light_pollution}. Therefore, optical stations cannot be operated under inappropriate weather conditions or during the daytime. 

\subsection{Radio observation methods}

To overcome these issues, alternative observation methods were developed. Apart form the infrasonic and seismic methods which are suitable for detection of large meteor impacts, there exist two radio waves observation methods which use the scattering ability of the ionized meteor trail \cite{infrasound}.

The oldest known radio method is a backscatter radar. It is an ordinary type of radar which expects meteor trails to be reflective targets. Currently, several radars of this type are in operation to study meteors SkiYMet \cite{skiymet}, CMOR \cite{CMOR_radar}.

However, all of these monostatic or bistatic radars have very small detection coverage, usually limited to the radar antenna field of view. Therefore this radar types observe only several spatially limited areas of the Earth's atmosphere. But the benefits of using this method are obvious - radio meteor detection capability is not dependent on the current weather and can work even during the daylight or nights with full Moon \cite{daylight_shover}.

Besides the above-mentioned scientific radar systems a multistatic radio meteor detection networks have evolved \cite{BRAMS}.
These forward scattering multistatic systems have great advantage of a large detection coverage. Unfortunately, the current spread of this technology is not sufficient to entirely cover the meteor flux of the Earth's atmosphere.
Therefore a new optimized and low cost Bolidozor network was developed. Bolidozor uses multistatic forward scattering approach which allows an efficient use of the radar transmitter energy  to maximize the information value collected from the meteor reflection.

\section{Forward scattering Method}

\begin{figure}
 \begin{center}
 \includegraphics[width=\linewidth]{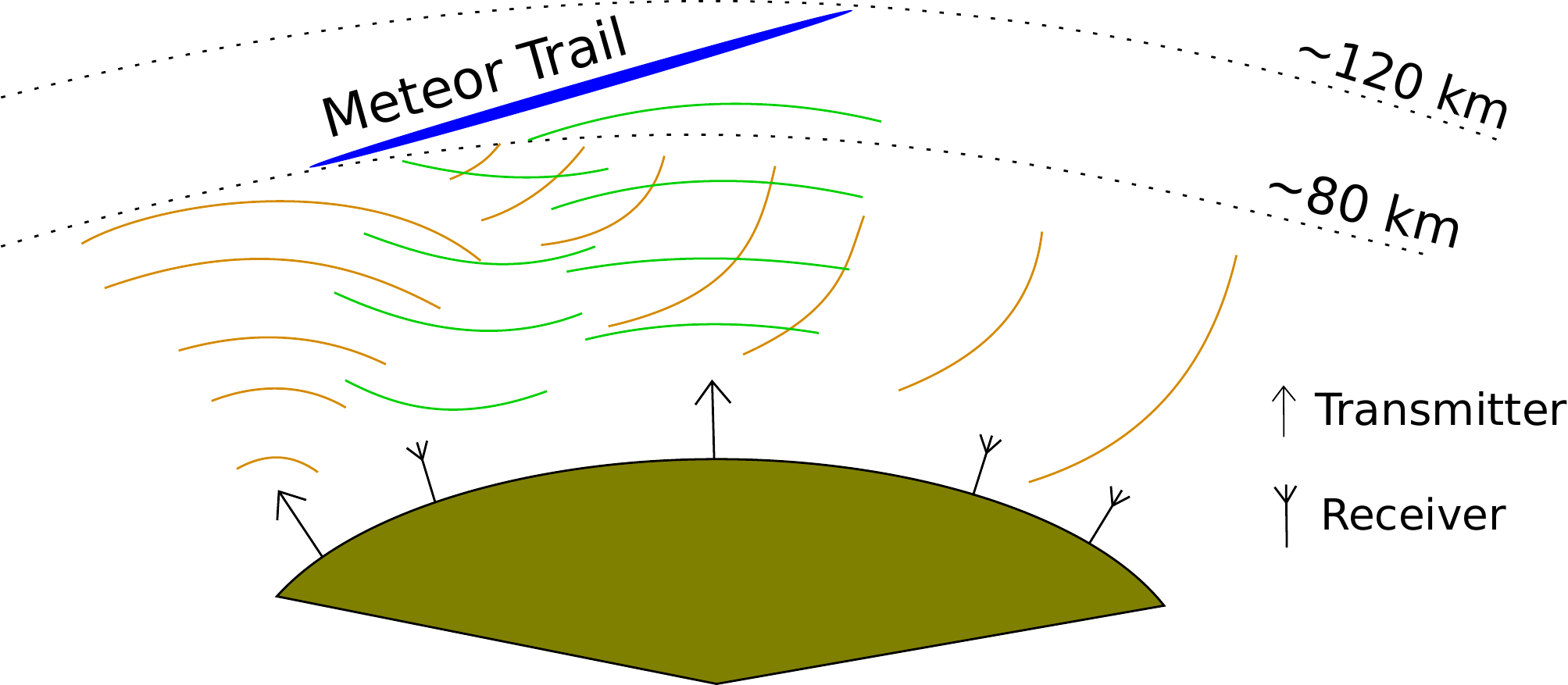}
 \caption{The method of radio meteor detection based on the forward scattering radar system}
  \label{fig:forward_scattering} 
 \end{center}
\end{figure}

The general principle of meteor observing by the forward scattering of radio waves off their trails is illustrated in figure \ref{fig:forward_scattering}. A radio receiver with operating frequency range of 30-200MHz is located at proper distance (about 500-2000 km) from a transmitter. A curvature of the Earth over this distance ensures no possibility of direct radio wave contact. When a meteoroid enters the atmosphere, its meteor trail may reflect the radio waves emitted by transmitter to the receiver. The signal can be received until the ionized meteor trail recombines. Reflections can last from a tenths of a second to a few minutes, depending on their frequency and ionization intensity. The received signal characteristics are related to physical parameters of the meteoric event \cite{forward_scatter}.

Bolidozor network currently uses GRAVES \cite{GRAVES_radar} transmitter located in France, which transmits a CW (continuous wave) signal at a frequency of 143.05 MHz. A radiation of the transmitter is directed mainly to the south hemisphere, but due to the imperfections of its antenna system, the signal from meteor reflections can be observed in almost all European countries. Therefore the transmitter is suitable for receivers' network operations with the aim of meteor trail detection and the development of algorithms for the calculation of meteor trajectory.

\begin{figure}
 \begin{center}
 \includegraphics[width=\linewidth]{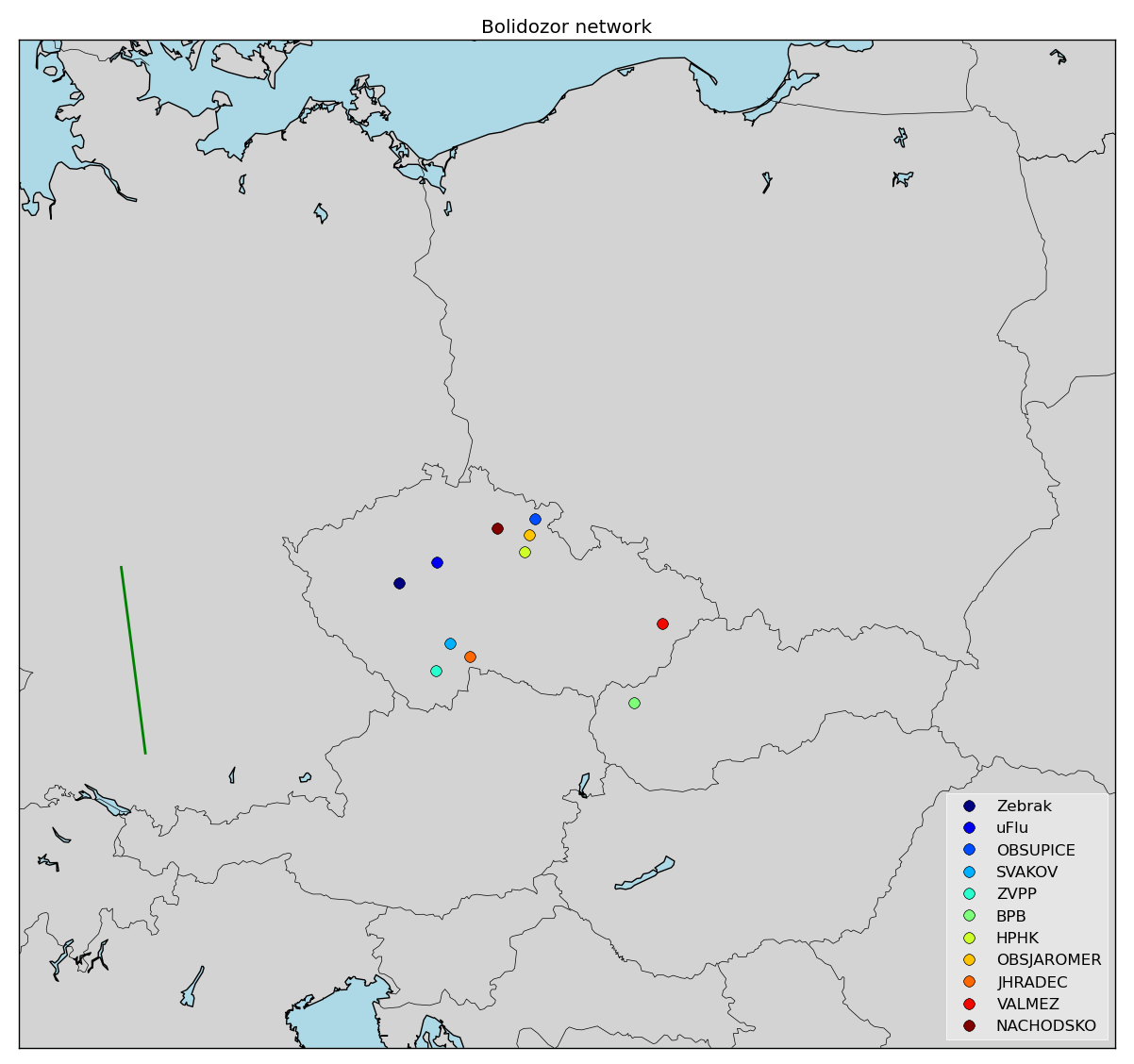}
 \caption{Bolidozor stations network}
  \label{fig:stanice_mapa} 
 \end{center}
\end{figure}
                   
The use of such high frequency beacon has one main advantage over the previous experiments.
Previous attempts used longer wavelengths in frequency range 20-50 MHz. Such long wavelengths were used to obtain a higher sensitivity to finer meteor trails. According to a simplified formula (\ref{equ:decay}), where $T$ is exponential time constant and $D$ is ambipolar diffusion coefficient and $\lambda$ is wavelength \cite{Decay_time}, we have longer meteor echos from meteors with the same ionisation energy observed by longer wavelengths compared to meteor reflections observed by shorter wavelengths. But shorter wavelengths allow us to detect finer details of meteor trails.

\begin{equation}
T = \frac{\lambda^2}{16 \pi ^2 D}
\label{equ:decay}
\end{equation}

\begin{figure*}
 \begin{center}
 \includegraphics[width=\textwidth]{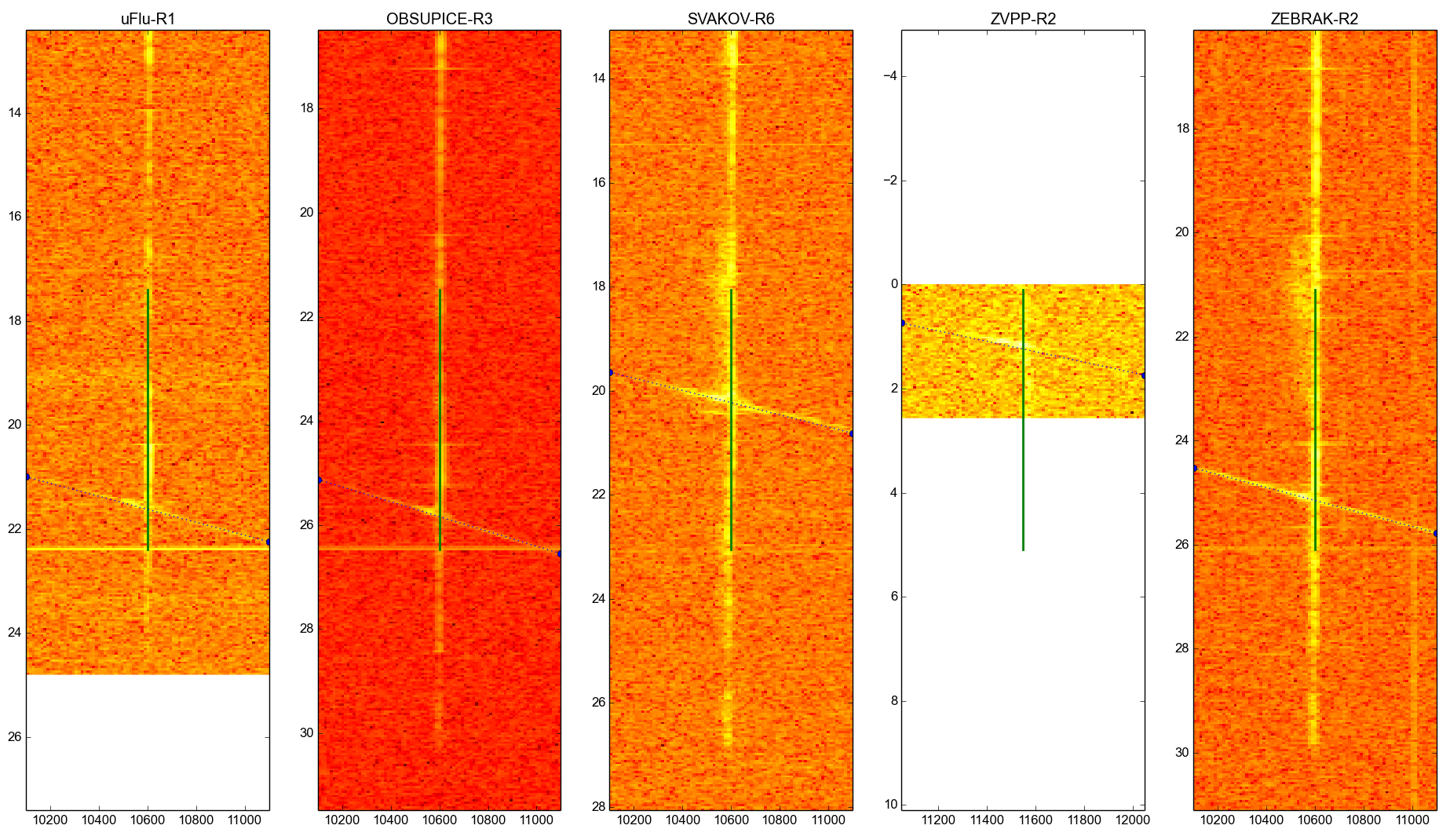}
 \caption{Example of meteor reflections for multiple stations - Spectrograms show the time aligned signal evolution over vertical axis. (The oldest data being at the bottom) Horizontal axis corresponds to frequency. (Highest frequency on right)}
  \label{fig:meteor_reflections} 
 \end{center}
\end{figure*}

The head echo of meteor is usually called as overdense due to its relatively high plasma frequency compared to used observation frequency ($F_{obs}$) in the front of the meteoroid shock wave that is created in the air. This condition is expressed by the equation (\ref{equ:plasma_frequency}). However, if we use a frequency near to the plasma frequency ($f_{pe}$) of the meteor trail we can distinguish the head echo and meteor trail reflection because the Doppler shift is applied on the part of reflected signal. This situation is shown in the figure \ref{fig:meteor_reflections} where head echoes are marked by sloped dotted lines. Static meteor trail reflections are marked by straight vertical lines.

\begin{equation}
F_{obs} << f_{pe} =\frac{\sqrt{\frac{n_e e^2}{m \epsilon_0}}}{2 \pi}
\label{equ:plasma_frequency}
\end{equation}
\begin{itemize}
\item $n_e$ is the number density of electrons,
\item $e$ is the elementary electrical charge,
\item $m$ is the effective mass of the electron,
\item $\epsilon_0$ is the permittivity of vacuum.
\end{itemize}

Obviously not all parts of the meteor trail are observable from one station as the signal can be scattered to different directions non-specularly.
But if we use multiple receiving stations, we greatly increase the statistical sensitivity because several stations could be located on the reflection spots. Therefore it is very useful to have stations in a form of cooperative detection network.
The network of stations has many advantages over a single transmitter - single receiver configuration. For example it brings robustness which allows operability even in case that a part of the system is under a maintenance and therefore not functional.
                   
There also exist signal processing advantages, especially if we want to compute meteor parameters such as its velocity and trajectory from the meteor radio observation. All physical parameters we could determine from the single reflection are signal intensity and frequency shift in a given time which corresponds to an ionisation intensity and bi-static velocity.
Therefore we must combine information about one meteor event from multiple stations to obtain points in space corresponding to the meteor trajectory. We should alight the events according to time stamps. Therefore, a precise time synchronization between stations is required. The exact required precision depends on the network geometry. But if we want to work in 300 m distance resolution, which is  typical for the current optical methods, we need the time precision on approximately the microseconds scale.

\section{Network Architecture}

\begin{figure*}
 \begin{center}
 \includegraphics[width=\textwidth]{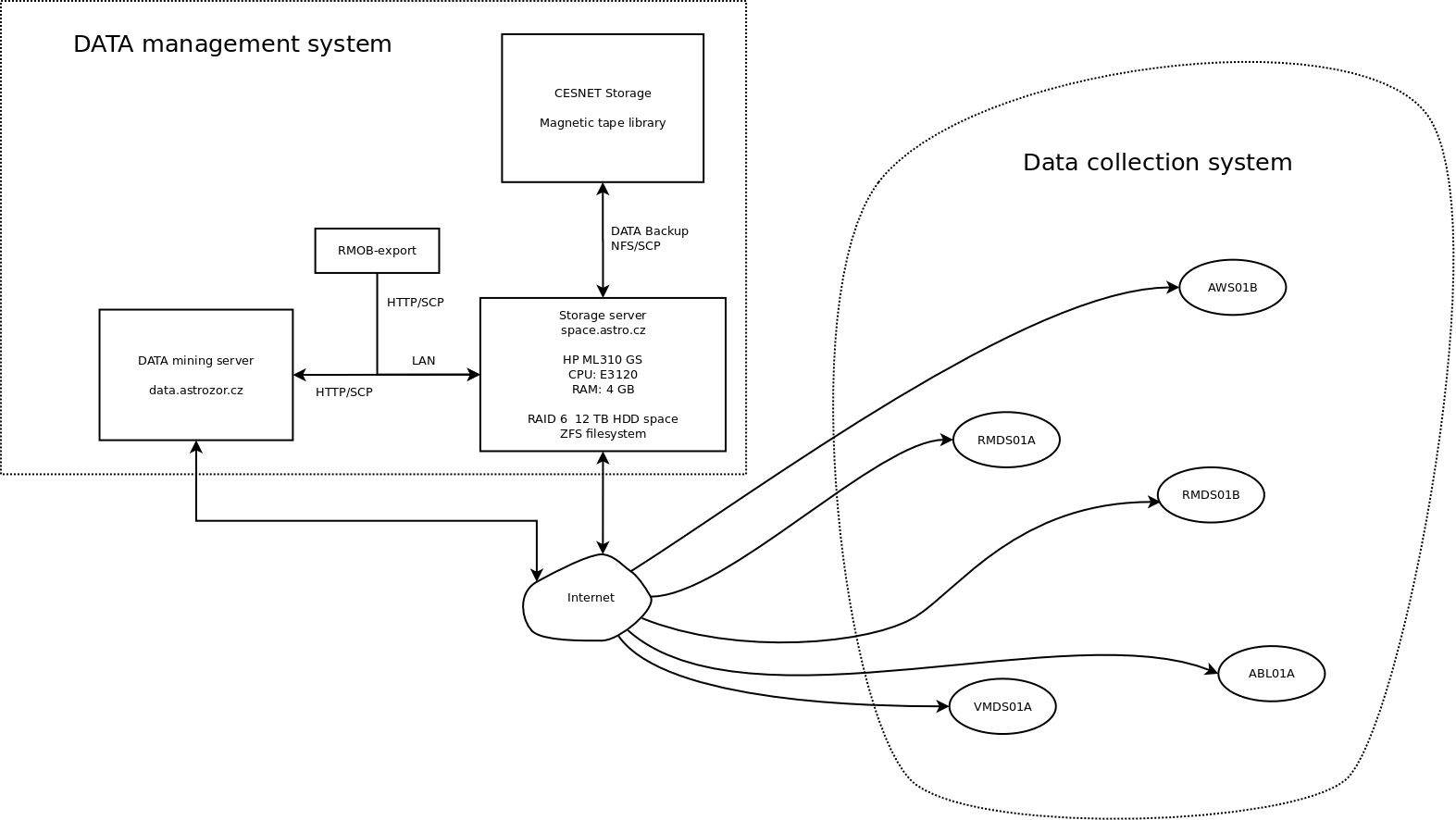}
 \caption{Bolidozor network interconnection schema.}
  \label{fig:Network_interconnection} 
 \end{center}
\end{figure*}

Bolidozor network consists of several types of stations: the reception stations, data servers and processing stations. Interconnection of these network nodes is displayed in figure  \ref{fig:Network_interconnection}.

Network data are generated exclusively by Radio Meteor Detection Station (RMDS) \cite{RMDS}  with current version RMDS02D and several other network data source stations are planed (Video meteor stations, infrasonic station etc.). 

Real-time transfer of all data volume is impractical because of stations' internet connection bandwidth limits and extensive data server load. Moreover, the stations are usually situated in remote sites or observatories without fast internet connection. Therefore only  the meteor events are recorded and stored locally on station's SDcard storage media. This design minimizes the network connectivity bandwidth demands and connection reliability.

The storage is large enough to handle several days of data records. The data from station storage is periodically transferd to the central data storage server (space.astro.cz). Process is controlled by standard utility Rsync which is commanded by python script.

The central data server currently contains around 7 TB of data (since 2014) and increases in size by approximately 1GB of data per station per day \cite{CAS}. All 9 million of data records are available for public via the HTTP requests. The server is backuped by CESNET Storage Department. 

Current centralized storage system is not optimal because expanding the storage capacity is quite complicated and expensive. Another decentralised approach based on purely data storage stations connected to the network was considered, but temporary rejected for its system's complexity and current unavailability of suitable tools. 

Stored data are processed by several separate data processing servers. Statistical processing is continuously performed on the stored data using the following technologies and programming languages: R, RapidMiner, Python \cite{iPython} and Jupyter. One example of an existing data processing facility is the system meteor1.astrozor.cz which downloads metadata files, searches for meteor events and prepares the outputs for rmob.org site.
Data processing system is continuously improved. The first implementation used was RapidMiner framework, but recent advantages in open-source scientific tools development allow us to use more sophisticated technologies like Jupyter \cite{Jupyter}. More interactive computational tools like \cite{scipy} emerged recently.

Several tools are prepared for simplification of potential data retrieving from the storage server. For example a python module bzbrowser was created. This module helps with the effective data downloading from the server so that anyone can use Bolidozor data for data mining.

\subsection{Detection Station Software}

Each detection node runs our software configuration on ARM based computer with Linux OS. The signal from receiver is captured by a modified sdr-widget server from GHPSDR3 project \cite{ghpsdr3}. SDR-widget server accepts socket connections from other processing and displaying software clients. The main software component is radio-observer code which detects the meteor events in data stream and creates data records as FITS files.

\begin{figure}
 \begin{center}
 \includegraphics[width=\linewidth]{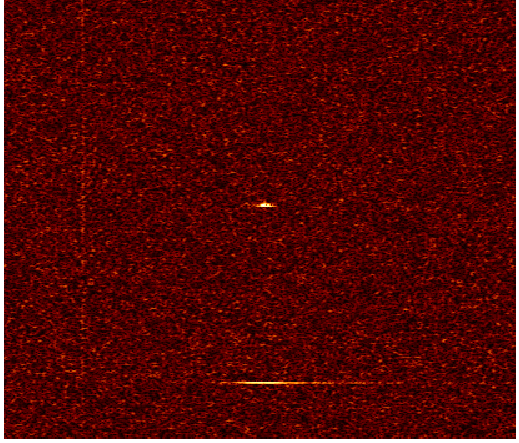}
 \caption{Example of snapshot spectrogram with several meteor reflections. (Vertical axis correnponds to time (60 s range), horizontal axis to frequency (600 Hz range)}
  \label{fig:FITS_snapshot} 
 \end{center}
\end{figure}

We are currently using several types of FITS recordings. 
\begin{itemize}
\item A 'snapshot' format continually records spectrograms of every minute. These spectrograms record only a small part of spectra where meteor echoes usually appear - an example of snapshot record is shown in figure \ref{fig:FITS_snapshot}. The purpose of snapshots is an illustrative recording of events which are not detected by radio-observer software. This could be useful for a new event type discovery or station debugging. e.g. if noise interference appears.
\item A more detailed format is represented by a 'met FITS' which contains spectrogram of the meteor reflections. The most detailed output file is the "raw FITS" which contains signal samples directly from the analog to digital converter. The RAW FITS is intended primarily for post-processing.
\end{itemize}

All FITS files contain metadata information in FITS's header, but for effective searching in data structure two metadata files are created in CSV format. First is meta.csv which contains information about events detected by radio-observer software e.g. meteor echo duration, frequency of maximal energy etc. Another output is freq.csv which contains recordings about local oscillator drifts and its corrections.

\subsection{Trajectory estimation possibility}

\begin{figure*}
 \begin{center}
 \includegraphics[width=\textwidth]{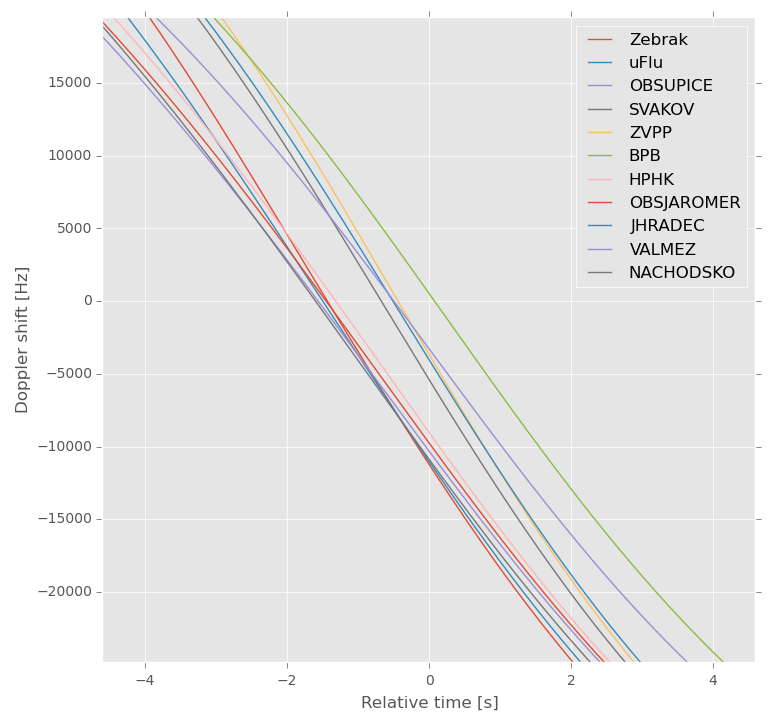}
 \caption{Doppler shifts calculated for meteor ground path displayed in the figure \ref{fig:stanice_mapa}.}
  \label{fig:dopplers} 
 \end{center}
\end{figure*}

Every meteor trajectory in atmosphere create its own Doppler shift footprint.  This process could be easily modelled by a numerical calculation of Doppler shifts in points of trajectory. For a simplicity, the presented model expects constant velocity along a straight line of the meteor path which is divided to equidistant time samples. A numerical difference of path distances between transmitter, meteor and receiver is calculated. Then velocity and Doppler shift value is obtained for every point of trajectory. 
The resulting figure of Doppler shifts calculated along the model meteor path visible to every existing station is shown in figure \ref{fig:dopplers}.

This signal model can be easily confirmed from meteor database where several meteor events are detected on multiple stations. If we plot such meteor event in time aligned spectrogram we obtain an image similar to the figure \ref{fig:meteor_reflections}.
Precise meteor trajectory estimation methods are intensively investigated at the moment.  One of the difficulties is a suboptimal geometry situation and low inter-station events correlation. Therefore expanding of the network is another task which runs in parallel.

\subsection{Station Hardware}

Bolidozor network main objective is a calculation of meteor trajectories from echoes. Therefore, we needed to develop a high performance receiver with unique parameters, especially with a high quality of time synchronization. Tight time synchronization requirements between nodes increase the complexity of the receiver system. To simplify the development process we used MLAB open source electronic prototyping platform. Therefore the whole station has an open hardware design \cite{MLAB}.

\begin{figure}
 \begin{center}
 \includegraphics[width=\linewidth]{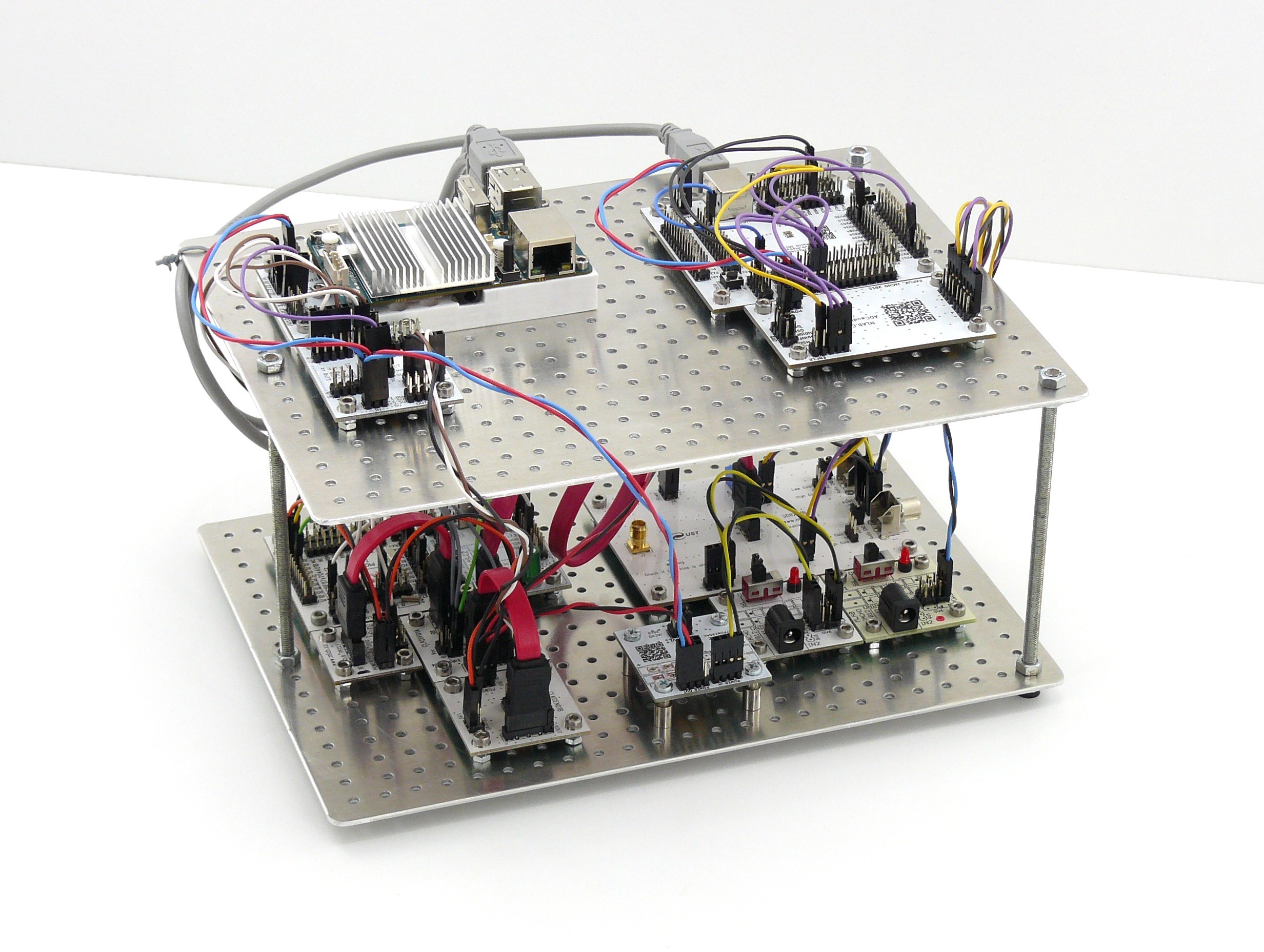}
 \caption{The Bolidozor detection station version RMDS02D.}
  \label{fig:RMDS02D} 
 \end{center}
\end{figure}

The use of modular concept allows us on-the-fly modifications and testing of new station's features. After verifying the new hardware features, other stations could be easily updated by adding or replacing MLAB modules. The current station hardware is shown in figure \ref{fig:RMDS02D}.

\begin{figure*}
 \begin{center}
 \includegraphics[width=\textwidth]{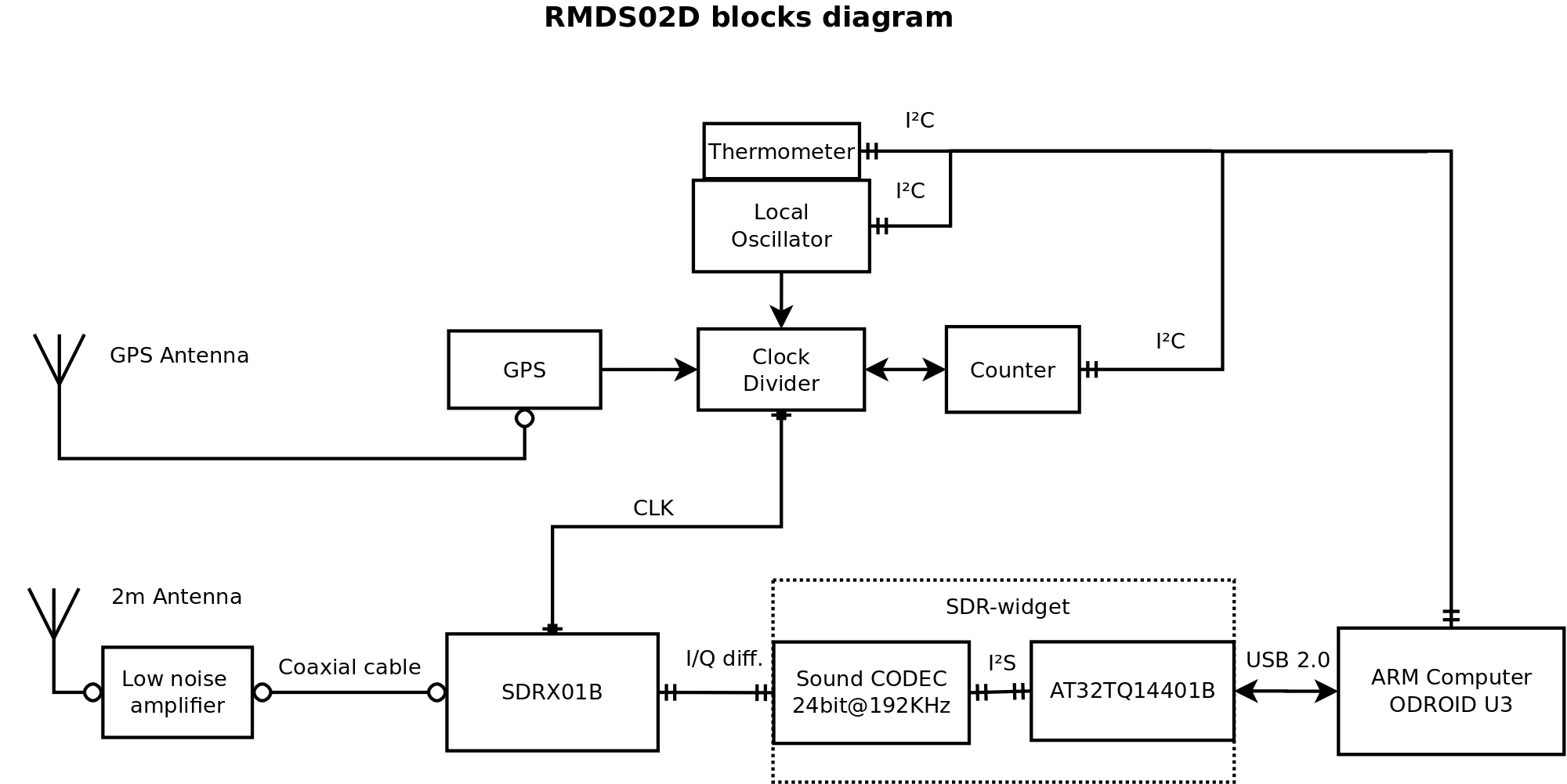}
 \caption{Block schematics of RMDS station.}
  \label{fig:block_schematics} 
 \end{center}
\end{figure*}

The station receiver consist of several subsystems shown in figure \ref{fig:block_schematics}.  At the input, the RF signal from a low noise amplifier placed near the antenna is conducted by a coaxial cable to the sampled mixer module. The mixer module is fed by the high stability local oscillator signal that is modulated by the time stamps. 
The local oscillator signal is synthesised by the TCXO (temperature compensated oscillator). Frequency errors of the oscillator are measured with the help of the GPS and used for a long term error compensation of the local oscillator frequency. Precise measurement of the oscillator frequency is executed by a frequency measurement circuit which completes the measurement once every ten seconds in absolute UTC time. 
Frequency measurement intervals of local oscillator are controlled by GPS PPS output. Measured oscillator parameters (frequency and temperature) are processed by a frequency-guard software running on the station computer. The software determines if some correction of a local oscillator is necessary and compensates the oscillator parameters. Therefore the resulting Allan variance of the local oscillator signal could be very low. 

The time marks in the meteor data are coded by PPS signal from GPS receiver which triggers a divider connected to the receiver and to the frequency counter, the clock signal to the signal mixer is interrupted by PPS impulses. Interruption of clock signal generates wide band spike which appears as horizontal line in spectrogram of the RF signal. Interruptions of signal could then be used as precise time stamps in post-processing algorithms. GPS Time marks linked to UTC have much better time resolution compared to the one used in current digitalization unit.  
    
Low frequency output of the mixer is directly digitized by SDR-widget compatible hardware \cite{SDR-widget}. The signal is currently digitized with 24bit resolution with sampling rate of 96 kHz. 
But parameters will be further improved by altering the station configuration. Actual precision are considered as adequate for a current stage of development process.

\subsection{Receiver parameters}

The minimum detectable signal at receiver input is approximately -120 dBm for 143 MHz with currently used software configuration where FFT resolution is around 4 Hz. However, we have placed a LNA01A low noise amplifier before the receiver directly at the antenna output. Gain of LNA01A is around 25 dB and noise figure of this amplifier is around 0.6 dB.  Such design allows us the use of small omnidirectional antennas for detection of lager meteors. We currently use a 1/4 lambda Ground Plane antenna for easy deployment to as many stations as possible, but another types of antennas with better parameters are currently in development process.

\section{Conclusion}

A new type of radio meteor detection system is being build, based on a new idea of distributed scientific measurement systems. At the moment Bolidozor network produces large volume of valuable radioastronomy data ready for further processing.  
We currently have a large database of multi-station meteor reflections. Unfortunately, we have yet not reached suitable network geometry to obtain meteor position and trajectory by the known methods \cite{Doppler_method}. 
To overcome this issue, the further evolution of Bolidozor network should be focused on the research of the new trajectory estimation algorithms. We are also working on the network extension with the aim of extension of the service area as well as increasing the stations density. 

The Bolidozor network depends primarily on volunteers that run and maintain the network nodes, therefore their work should be appreciated accordingly.
We would like to thank the Czech Astronomical society for maintenance of our central data server space.astro.cz
Thanks to Center of machine perception of Czech Technical University for financial support of Bolidozor system.

Thanks to SDR-widget development team for open source hardware and related software. The study is supported by the SGS grant No SGS16/166/OHK3/2T/13 of the Czech Technical University in Prague.

\end{document}